\documentclass[pre,amsmath,amssymb,twocolumn,superscriptaddress]{revtex4}
\usepackage{amsmath, amstext, amssymb, amsfonts, amsxtra}
\usepackage[usenames]{color}
\usepackage{grffile}
\newcommand{\camerino}{Department of Physics, University of Camerino, 61032 Camerino, Italy}
\newcommand{\tongji}{Center for Phononics and Thermal Energy Science, School of Physics 
Science and Engineering, Tongji University, Shanghai 200092, China}
\newcommand{\como}{Dipartimento di Scienza e Alta Tecnologia,
Center for Nonlinear and Complex Systems, Universit\`a degli Studi dell'Insubria,
via Valleggio 11, 22100 Como, Italy}
\newcommand{\infn}{Istituto Nazionale di Fisica Nucleare, Sezione di Milano,
via Celoria 16, 20133 Milano, Italy}
\newcommand{\brazil}{International Institute of Physics, Federal University
of Rio Grande do Norte, Campus Universit\'ario - Lagoa Nova, CP. 1613,
Natal, Rio Grande Do Norte 59078-970, Brazil}
\newcommand{\NEST}{NEST, Istituto Nanoscienze-CNR, I-56126 Pisa, Italy}
\newcommand{\xiamen}{Department of Physics and Key Laboratory of Low
Dimensional Condensed Matter Physics (Department of Education of Fujian
Province), Xiamen University, Xiamen 361005, Fujian, China}
\newcommand{\lanzhou}{Lanzhou Center for Theoretical Physics, Lanzhou
University, Lanzhou 730000, Gansu, China}

\begin{document}

\title{Autonomous circular heat engine}
%\title{A cycle thermoelectric engine model and its power-efficiency trade-off}

\author{Giuliano Benenti}
\affiliation{\como}
\affiliation{\infn}
\affiliation{\NEST}
\author{Giulio Casati}
\affiliation{\como}
\affiliation{\brazil}
\author{Fabio Marchesoni}
\affiliation{\camerino}
\affiliation{\tongji}
\author{Jiao Wang}
\affiliation{\xiamen}
\affiliation{\lanzhou}

\begin{abstract}
A dynamical model of a highly efficient heat engine is proposed, where an applied
temperature difference maintains the motion of particles around the circuit consisting 
of two asymmetric narrow channels, in one of which the current flows against the 
applied thermodynamic forces. Numerical simulations and linear-response analysis 
suggest that, in the absence of frictional losses, the Carnot efficiency can be 
achieved in the thermodynamic limit.
\end{abstract}

\maketitle

\section{Introduction}

In the past decades, studying classical and quantum transport from the
microscopic dynamics perspective has led to major advances in our understanding
of heat conduction in low-dimensional systems~\cite{PRL84, MP2000, PR377, AP57,
RMP84, Lepri, BCMP16, BLL21}, unveiling fundamental mechanisms of normal and
anomalous transport and the conditions for heat rectification. In more recent
years, investigations have been extended to more complicated situations
involving two or more coupled currents, like in thermoelectric~\cite{PR694}
or thermodiffusive~\cite{parola} transport. Also in this case, a microscopic
approach has played a unique role, finding new paths to achieve the Carnot
efficiency in heat-to-work conversion~\cite{PRL2018, PRL2013}, extending Onsager
reciprocal relations to systems with broken time-reversal symmetry~\cite{Rondoni14, 
PRR2020}, and discovering the highly counterintuitive phenomenon of inverse 
currents, whereby an induced current flows opposite to the applied thermodynamic forces~\cite{CMP18,ICC,Zhang21}.

We propose to extend these notions to mass transport in low dimensional geometries,
like narrow channels, where particles tend to move in single-files~\cite{Marchesoni}.
Indeed, when their diameter is comparable to the channel cross section, particles
either do not pass each other at all, or do so only by overcoming a repulsive
potential barrier and, possibly, at the cost of frictional losses~\cite{Mom2002}.
However, single files subject to thermal gradient not only pose a fundamental
problem for themselves (with potential applications to natural and artificial
nano-devices~\cite{Marchesoni}) but can also be suitably arranged to produce
coupled particle currents.

Here, we demonstrate the possibility of building a circular heat engine
consisting of two channels coupled to two particle reservoirs maintained 
at different temperatures. In addition, each channel contains particles of a
different species, which either repel or bypass the particle flowing between
the reservoirs, depending on their masses and velocities. For an appropriate
choice of these parameters, the particle current in one channel may flow from 
low to high temperatures, and a stationary circular current between the 
reservoirs is established.
As a consequence, the engine can convert a substantial fraction of the heat
flowing from the hot to the cold reservoir into work. Extensive numerical
simulations and a linear-response analysis suggest that the Carnot efficiency
can indeed be achieved in the thermodynamic limit, i.e., for infinitely
long channels.

\section{Model and engine mechanism}

\begin{figure}[!]
\includegraphics[width=8.8cm]{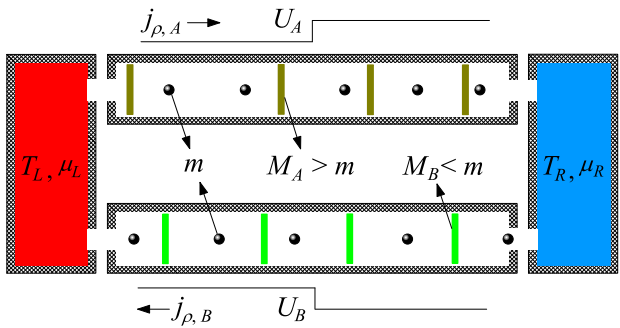}
\caption{Sketch of a two-channel engine working between two reservoirs of
fixed temperatures $T_L$ and $T_R$ and adjustable chemical potentials
$\mu_L$ and $\mu_R$. The two particle species in each channel are represented
by bullets of mass $m$ and rods of mass $M_{A,B}$. When a stationary circular
current of bullets is established, the engine can work against applied
potentials $U_{A,B}$. In this drawing, the channels have equal lengths,
$L_A=L_B$.}
\label{fig:model}
\vskip-.3cm
\end{figure}

A sketch of the proposed engine is
plotted in Fig.~\ref{fig:model}. It is made of two one-dimensional (1D)
channels of length $L_\Xi$ ($\Xi=A, B$), coupled at their end points to two
reservoirs of temperatures $T_k$ ($k=L, R$). The number of particles in the
overall system (channels plus reservoirs) is fixed. Each channel contains two
species of particles, graphically represented by bullets of mass $m$ and rods
of mass $M_\Xi$, respectively. Dynamics inside the channels is purely Hamiltonian.
All particles move freely, except when they collide with one another or hit a
channel end point. When two particles collide, they either pass through each
other, if their total energy in the frame of their center of mass is larger
than a fixed inner potential barrier $h_\Xi$ or simply bounce back; in both
cases the pair momentum and the pair kinetic energy are conserved. The particle 
dynamics adopted here is inspired to well-known models for single-file diffusion 
and granular fluids, where $h_\Xi$ can be regarded as the energy barriers associated 
with the configurational changes two particles undergo whereas passing each other in 
the channel~\cite{Marchesoni,Ruthven,Bukowski}. For the engine to work it is essential 
that the rods in the two channels are different. When a rod hits a channel-reservoir 
boundary, it is reflected back with a newly assigned velocity sampled from a certain 
distribution determined by the reservoir temperature (see below). Accordingly, the 
number of rods in each channel is conserved, as they only exchange energy with the 
reservoirs. As for the bullets,
when one reaches a channel boundary, it will enter the connected reservoir.
Meanwhile, the reservoirs keep injecting bullets into the channels with rates
and energy distributions determined by their temperatures $T_k$ and densities
(or chemical potentials $\mu_k$). Following these simple rules, after an
appropriate transient, a steady-state circular current of bullets sets in, 
sustained by the temperature difference imposed by the reservoirs; the bullet 
densities (and the chemical potentials $\mu_k$) in each (large but finite) 
reservoir autonomously adjust to support such a current. Ultimately, the 
circulating particle current and, thus, the possibility of extracting useful 
work depend on the fixed temperature difference between the reservoirs.

%%%%%%%%%%%%%%%%%
%%%%%%%%%%%%%%%%%

Suppose that $T_L>T_R$ and $\mu_L>\mu_R$. Intuitively, we may expect that in
each channel, both the energy and the bullet current would flow forward from
left to right. This is indeed the case when $M_\Xi>m$. In sharp contrast, for
$M_\Xi<m$, either the energy or the bullet current -- depending on the parameter
choice -- may flow on reverse from right to left, against both thermodynamical
forces~\cite{ICC}. The mechanism of current reversal is related to the fact that
under the given collision rules, the probability for a bullet-rod pair to cross
each other is higher when the light (heavy) particle is on the hot (cold) reservoir
side because in such a case, their relative velocities are likely to be higher.
The kinetics of bullets and rods, thus, causes a left-right unbalance in their
densities along the channel~\cite{SONPT}.
%As the temperature difference between the reservoirs
%is increased, this mechanism may even result in a phase separation~\cite{SONPT}.

Thanks to this peculiar dynamical effect, we can make the bullets flow on reverse
in one channel, say, channel $B$, by setting $M_B<m$  and, thus, create a clockwise
circular bullet current through the whole system. Such current can work also
against an external bias, represented by the potentials $U_{A,B}$ in
Fig.~\ref{fig:model}~\textcolor{blue}{\cite{potential}}.
For instance, one could use
part of the kinetic energy of the
circulating particles to lift a weight. A fraction of the heat flowing from the
hot to the cold reservoir would be then converted into mechanical work, the rest
being dumped into the cold reservoir. This fulfills the function of an engine.

\section{Linear response analysis}

Before presenting the output of our numerical simulations, we analyze the model
in the  linear-response regime. We adopt here for concreteness the language of 
thermoelectricity, with charged particles circulating, such as in a thermocouple, 
but our results apply equally well to other coupled flows, such as in thermodiffusion. 
We start from the linear transport equations~\cite{Callen, Groot} for channel 
$\Xi$ $(\Xi=A, B)$,
\begin{equation}
\left(
\begin{array}{c}
J^{\rho}_{\Xi}\\
J^{u}_{\Xi}
\end{array}
\right) = \left(
\begin{array}{cc}
\mathcal{L}^{\rho \rho}_{\Xi} & \mathcal{L}^{\rho u}_{\Xi} \\
\mathcal{L}^{u \rho}_{\Xi} & \mathcal{L}^{u u}_{\Xi}
\end{array}
\right) \left(
\begin{array}{c}
\mathcal{F}^{\rho}_{\Xi}/L_\Xi\\
\mathcal{F}^{u} /L_\Xi
\end{array}
\right).
\label{eq:JrJu}
\end{equation}
Here $J^{\rho}_{\Xi}$ and $J^{u}_{\Xi}$ are the particle (bullet) and the energy
currents, $\mathbb{L}_\Xi=(\mathcal{L}^{ij}_{\Xi}$) ($i,j=\rho, u$) the matrix of
the Onsager kinetic coefficients, and $\mathcal{F}^{\rho}_{\Xi}$ and $\mathcal{F}^{u}$
the thermodynamical forces, defined as
$\mathcal{F}^{\rho}_A=\mu_L \beta_L-(\mu_R+U_A) \beta_R$,
$\mathcal{F}^{\rho}_B=\mu_L \beta_L-(\mu_R-U_B) \beta_R$,
and
$\mathcal{F}^{u}=\beta_R-\beta_L$, respectively, with $\beta_k=1/(k_B T_k)$
($k=L, R$, $k_B$ is the Boltzmann constant).
The Onsager coefficients are related with the familiar transport coefficients, i.e.,
the electrical conductivity $\sigma_\Xi$, the thermal conductivity $\kappa_\Xi$, and
the thermopower $S_\Xi$, as follows:
\begin{equation}
\sigma_\Xi=\frac{e^2}{T}\,L^{\rho\rho}_\Xi,
~\kappa_\Xi=\frac{1}{T^2}\frac{\det\mathbb{L}_\Xi}{L^{\rho\rho}_\Xi},
~ S_\Xi=\frac{1}{eT}\left(\frac{L^{\rho u}_\Xi}{L^{\rho\rho}_\Xi}-\mu\right),
\label{eq:coeff}
%\nonumber
\end{equation}
where $e$ is the charge of each particle, $T\approx T_L\approx T_R$, 
$\mu\approx \mu_L\approx \mu_R$ in linear response
approximation. We can then rewrite Eq.~(\ref{eq:JrJu}) as
\begin{equation}
\left\{
\begin{array}{l}
L_A J^u_A=\kappa_A^\prime\Delta T + T \sigma_A S_A (\Delta \mu - U_A),\\
L_B J^u_B=\kappa_B^\prime\Delta T + T \sigma_B S_B (\Delta \mu + U_B),\\
L_A J^\rho_A=\sigma_A S_A \Delta T + \sigma_A (\Delta \mu - U_A),\\
L_B J^\rho_B=\sigma_B S_B \Delta T + \sigma_B (\Delta \mu + U_B).
\end{array}
\right.
\nonumber
\label{eq:currents}
\end{equation}
Here $\kappa_A^\prime=\kappa_A+ T \sigma_A S_A^2$,
$\kappa_B^\prime=\kappa_B+ T \sigma_B S_B^2$,
$\Delta T=T_L-T_R$, and $\Delta\mu=\mu_L-\mu_R$. On imposing the circular,
steady-flow condition
\begin{equation}
J^\rho_A+J^\rho_B=0,
\label{eq:SF}
\end{equation}
the output power $P$ and the efficiency $\eta$ read respectively
\begin{equation}
P=J^\rho_A U_A-J^\rho_B U_B, ~~~~~ \eta={P}/(J^u_A+J^u_B).
\label{eq:efficiency}
\end{equation}
Note that by using the steady-flow condition, $P$ and $\eta$ can be explicitly 
rewritten as functions of $U_A+U_B$, rather than of $U_A$ and $U_B$, separately 
(see Appendix~\ref{appA}).
Moreover, both the maximum efficiency $\eta_{\rm max}$ and the
efficiency at the maximum power $\eta(P_{\rm max})$ have the usual
dependence~\cite{PR694} on a nondimensional figure of merit,
\begin{equation}
YT=\frac{(\sigma_A/L_A)(\sigma_B/L_B)(S_A-S_B)^2}{(\sigma_A/L_A+\sigma_B/L_B)
(\kappa_A/L_A+\kappa_B/L_B)}\,T,
\label{eq:YT}
\end{equation}
(in lieu of the thermoelectric figure of merit $ZT$~\cite{Prosen09, PRE2009}), namely
\begin{equation}
\eta_{\rm max}=\eta_C
\frac{\sqrt{YT+1}-1}{\sqrt{YT+1}+1}, \,\,\,
\eta(P_{\rm  max})=\frac{\eta_C}{2}\frac{YT}{YT+2},
\label{eq:eta}
\end{equation}
with Carnot efficiency $\eta_C=1-T_R/T_L$ (we assume $T_L>T_R$) and maximum power
\begin{equation}
P_{\rm max}=\frac{1}{4}\,\frac{\sigma_A\sigma_B}
{\sigma_AL_B+\sigma_BL_A}\,(S_A-S_B)^2(\Delta T)^2.
\nonumber
\end{equation}
The power-efficiency trade-off for a given value of $YT$ can also be obtained
(see, e.g., Ref.~\cite{PR694}),
\begin{equation}
\frac{\eta}{\eta_{C}}=
\frac{P/P_{\rm max}}{
2[1+2/(YT)\mp \sqrt{
1-P/P_{\rm max}}~]}.
\label{eq:loop}
\end{equation}

A limiting case is represented by the conventional thermocouple
configuration with $L_A=L_B\equiv L$, $\sigma_A=\sigma_B\equiv \sigma$,
$\kappa_A=\kappa_B\equiv \kappa$, and $S_A=-S_B\equiv S$. Here, the
figure of merit $ZT$ is recovered, $ZT=YT$, and the maximum power,
$P_{\rm max}=(\sigma/2L)S^2(\Delta T)^2$, amounts to twice the
maximum power of a single channel.

\section{Numerical study}

Our numerical results show that the proposed engine is very efficient.
In our simulations, we model the reservoirs as 1D ideal gases of
bullets~\cite{reservoir}. They inject bullets into the channels randomly
in time, with constant rates $\gamma_k$ ($k=L, R$)~\cite{rate},
$\gamma_{k}=(\rho_0/\sqrt{2\pi m \beta_0})
({\beta_0}/{\beta_k})e^{\mu_k \beta_k-\mu_0 \beta_0}$,
where $T_0=1/(k_B \beta_0)$, $\rho_0$, and $\mu_0$ are, respectively, the
temperature, particle number density, and chemical potential of a reference
state (see below). The injection intervals thus follow the Poisson distribution
$\pi_k (t)=\gamma_k e^{-\gamma_k t}$, whereas the speeds of the injected particles
are sampled according to the Maxwell distribution~\cite{bath},
$P_k(v,m)={m|v|\beta_k}e^{-m v^2\beta_k/2}$.
Accordingly, when a rod particle of mass $M_\Xi$ hits the channel boundary next
to reservoir $k$, it bounces back with speed distribution $P_{k}(v, M_\Xi)$.

To establish the circular, steady-flow condition of Eq.~(\ref{eq:SF}), rather
than simulating the entire closed system (that is, channels and reservoirs),
we simulated first the two channels, separately, and computed the two curves
$J^{\rho}_{\Xi}$ vs $U_\Xi$  ($\Xi=A, B$). Then, for a given value of
$J^{\rho}\equiv J^{\rho}_A =-J^{\rho}_B$, we determined the corresponding values
of $U_{A,B}$, to be used in the subsequent simulation steps, which involve both
channels simultaneously. After the system had relaxed to its stationary state,
we computed the time-averaged currents and from Eqs.(\ref{eq:efficiency}),
the relevant output power $P$ and the efficiency $\eta$.

In the simulations reported here, we set $T_L=T+\Delta T/2$,
$\mu_L=\mu+\Delta \mu/2$, $T_R=T-\Delta T/2$, and $\mu_R=\mu-\Delta \mu/2$.
The number of rods in each channel is set to half of the expected particle
number of a 1D ideal gas at the equilibrium with the assigned $T$ and $\mu$,
i.e., $N_{rod, \Xi} = \rho L_\Xi/2$  with
$\rho=\rho_0 \sqrt{\beta_0/\beta} e^{\beta\mu-\beta_0\mu_0}$.
As for the reference state, we set $\rho_0=1$, $T_0=1$, and $\mu_0=0$
(in units such that $m=1$, $e=1$, and $k_B=1$). To make the system evolve in time, 
we implemented an effective event-driven algorithm~\cite{algorithm}, which yields 
output data points with relative error less than 0.5\%.

%%%%%%%%%%%%%%%%%
\begin{figure}[!]
\includegraphics[width=8.5cm]{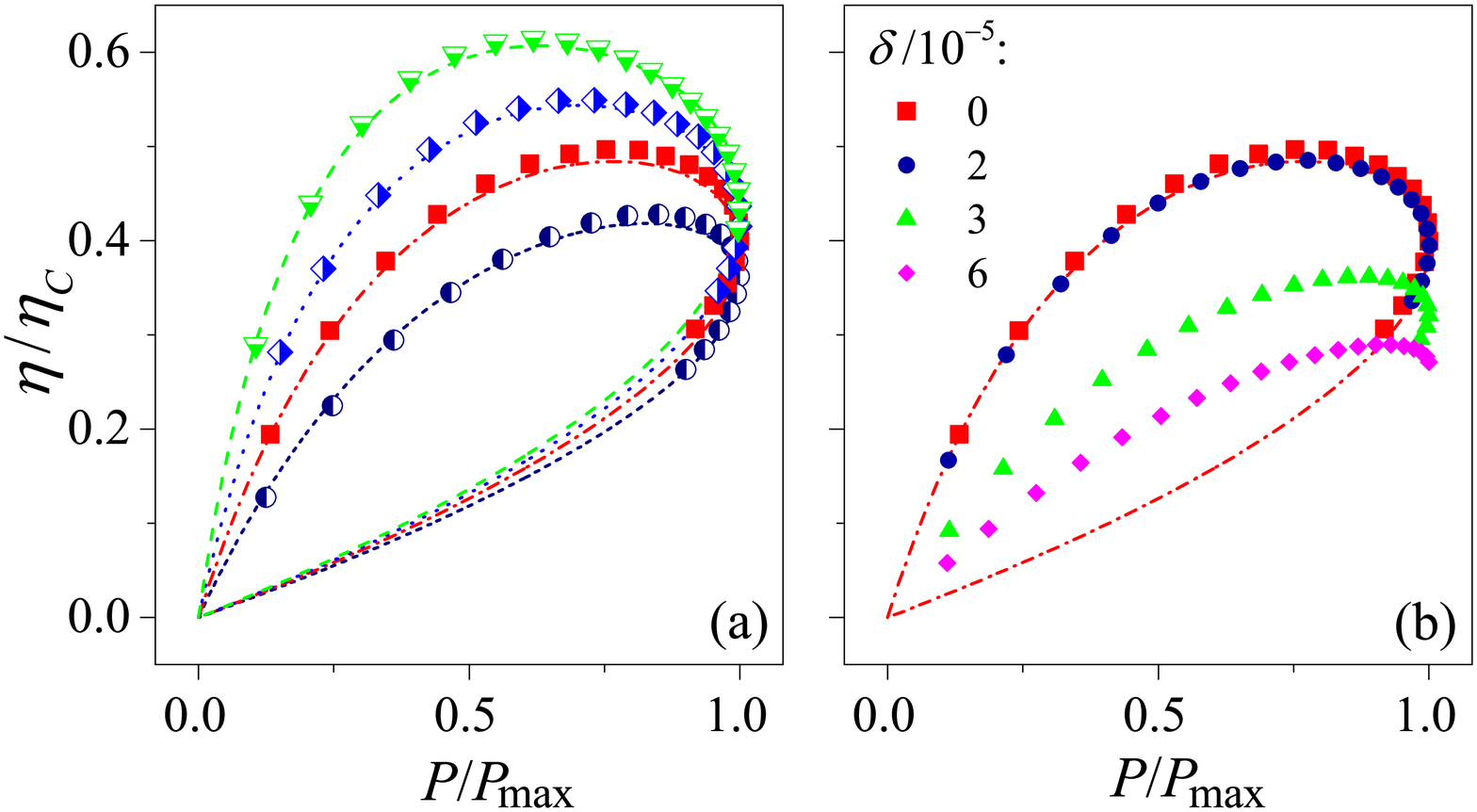}
\caption{(a) Efficiency, $\eta$, vs output power $P$ for two channels
of equal length $L_A=L_B$. Blue dots, red squares, blue diamonds, and green
triangles are for $L_\Xi=100$, $200$, $400$, and $1600$, respectively. The
curves next to each data set are obtained via Eq.~(\ref{eq:loop}) with numerically
computed $YT$. Fixed model parameters are as follows: $h_A=h_B=1$, $M_A=1.5$, $M_B=0.5$,
$T=1$, and $\mu=1.5$; Other tunable parameters are $\Delta T=0.1$ and $\Delta \mu=0.15$. 
(b) $\eta$ vs $P$, such as in (a), for $L_\Xi=200$ but finite dissipation parameter 
$\delta$ (see text below).}
\label{fig:loops}
\end{figure}
%%%%%%%%%%%%%%%%%

Our data show that this model works as an autonomous engine in a wide range
of parameters. Typical numerical results for the efficiency and the power
are displayed in Fig.~\ref{fig:loops}(a), where for any given system size
$L_A =L_B$, a data point represents the result obtained for a certain value
of $J^{\rho}$ %= J^{\rho}_A=-J^{\rho}_B$
(or $U_A$ and $U_B$). The closed curve
next to each set of the data points is the prediction of Eq.~(\ref{eq:loop}),
plotted for the corresponding value of $YT$, also obtained by numerical simulation
as explained below. The linear-response theory reproduces quite closely the numerical
data of Fig.~\ref{fig:loops}(a). Moreover, such agreement improves with increasing the
system size as expected since the temperature gradient $(\nabla T)_\Xi=\Delta T/L_\Xi$,
decreases upon increasing $L_\Xi$ at constant $\Delta T$. More importantly, we notice
that as $L_\Xi$ increases, the efficiency versus power curves shift upwards, meaning
that the engine performance improves. This remark hints at the possibility that the
figure of merit $YT$ in Eq.~(\ref{eq:loop}) is a monotonically increasing function
of the system size.

\begin{figure}[!t]
\vskip-0.1cm
\includegraphics[width=8.8cm]{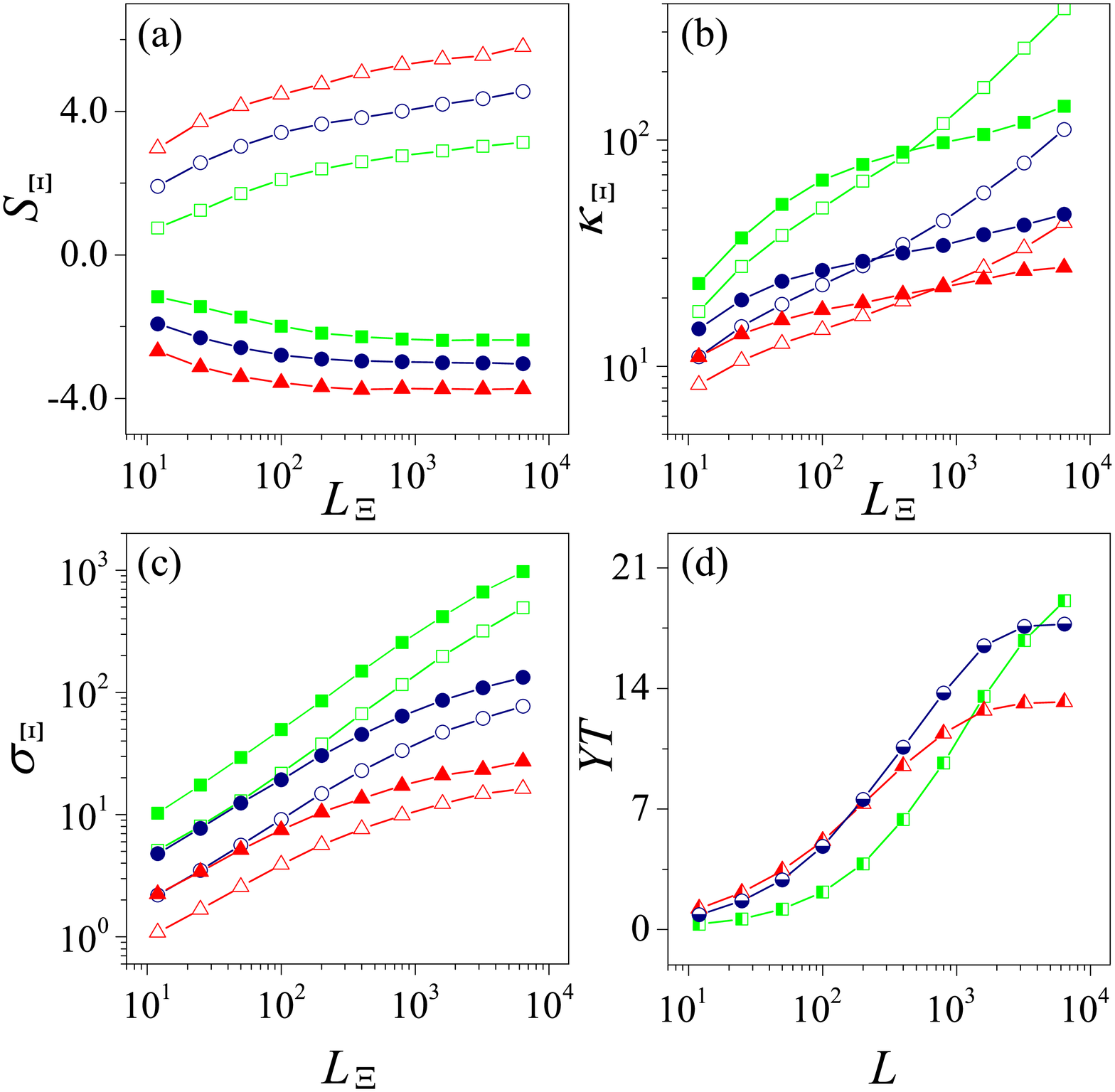}
\caption{(a)-(c) Transport coefficients of the two channels vs their lengths,
where the empty and the full symbols are for channel $A$ and channel $B$, respectively.
(d) $YT$ vs $L_A=L_B\equiv L$ for $h_A=h_B$. In all four panels, green squares, blue dots,
and red triangles are for, respectively, $h_\Xi =0.5$, $1.0$, and $1.5$, %($\Xi = A,B$).
All other simulation parameters are the same as in Fig.~\ref{fig:loops}(a).}
\label{fig:sgm-S-k}
\end{figure}

To determine $YT$ we made use of Eq.~(\ref{eq:YT}), where the transport
coefficients are also to be computed numerically. To this purpose, the two channels
were considered separately. For each channel, we followed the method detailed in
Ref.~\cite{rate}, i.e., the particle and energy currents were measured twice,
namely, for $\mathcal{F}^{\rho}_\Xi \ne 0$ and $\mathcal{F}^{u} =0$, and for 
$\mathcal{F}^{\rho}_\Xi =0$ and $\mathcal{F}^{u}\ne 0$ (having set $U_\Xi=0$).
The Onsager kinetic coefficients can then be evaluated through Eq.~(\ref{eq:JrJu}).
The corresponding transport coefficients, Eq. (\ref{eq:coeff}), computed for three
different potential barrier values, $h_A=h_B=0.5$, $1.0$, and $1.5$, are displayed
in Figs.~\ref{fig:sgm-S-k}(a)-\ref{fig:sgm-S-k}(c) vs $L_\Xi$. First of all, we note 
that due
to the inverse bullet current in channel $B$, the off-diagonal elements of the Onsager
matrix are negative~\cite{Lepri2012,ICC}. Accordingly, the Seebeck coefficients $S_A$
and $S_B$ are, respectively, positive and negative [see Fig.~\ref{fig:sgm-S-k}(a)],
which enhances the figure of merit $YT$ through the quadratic factor $(S_A-S_B)^2$
in Eq.~(\ref{eq:YT}). This is the key advantage of our model. Moreover, due to the
fact that the momentum is the only conserved mechanical quantity both $S_\Xi$ and
factor $(S_A-S_B)^2$ in our expression for $YT$ are predicted to saturate in
the thermodynamic limit~\cite{PRL2013}, a prediction corroborated by our numerical
simulations.

%%%%%%%%%%%%%%%%%
%%%%%%%%%%%%%%%%%

As for the electrical and thermal conductivities, it helps consider two limits for
the potential barriers $h_\Xi$. In the limit $h_\Xi \to \infty$, all particles turn
out to be hard core and the two species in each channel are always nonpassing. The
bullet currents are therefore blocked, thus resulting in $\sigma_\Xi = 0$. On the
other hand, according to the 1D heat conduction theory~\cite{PR377, AP57, Beijeren,
Spohn}, in the thermodynamical limit the heat conductivity would diverge, such as
$\kappa_\Xi \sim L_\Xi^{\nu}$ with $0<{\nu}<1$. Consequently, see Eq.~(\ref{eq:YT}),
in the same limit both $YT$ and the engine efficiency would vanish. In the opposite
limit $h_\Xi \to 0$ the dynamics becomes integrable and the bullets flow through 
the channel freely; hence, $\sigma_\Xi \sim L_\Xi$ and $\kappa_\Xi \sim L_\Xi$.
This working regime is not advantageous, either. Indeed, the bullet-rod
interaction is crucial to maintain an inverse current~\cite{ICC} in channel $B$.
Therefore, when the bullet-rod interaction vanishes for $h_B\to 0$, so does the
inverse current. More precisely, there exists a critical value $h_B^*$, such that
for $h_B < h_B^*$, the bullet current in channel $B$ starts flowing forward and our
engine, thus, stops working. Note that $h_B^*\sim L_B^{-0.5}$ (see Appendix~\ref{appB}) 
so that $h_B^*\to 0$ as $L_B\to \infty$. It is, therefore, reasonable to conjecture that 
for a given system size, one can determine the optimal finite values of $h_{\Xi}$ that 
maximize $YT$.

Based on the data in Figs.~\ref{fig:sgm-S-k}(a)-~\ref{fig:sgm-S-k}(c), we can 
investigate how $YT$
depends on the system size for assigned values of $h_\Xi$. We find that, in general,
when $L_A+L_B$ is fixed, $YT$ reaches its maximum for $L_A\simeq L_B$ and $h_A=h_B$.
For this reason, we focus on cases with $L_A=L_B$ and $h_A=h_B$ and in
Fig.~\ref{fig:sgm-S-k}(d) plot $YT$ vs $L_\Xi$ for three values of $h_\Xi$. $YT$
is confirmed to be an increasing function of $L_\Xi$, which saturates asymptotically
to a value that increases as $h_\Xi$ decreases. In addition, the optimal $h_\Xi$
value that, for a given $L_\Xi$, maximizes $YT$ decreases as $L_\Xi$ increases 
[~from Fig.~\ref{fig:sgm-S-k}(d),
we can tell that the optimal $h_\Xi$ value is larger than one for $L_\Xi<150$ but
smaller than one for $L_\Xi>4000$]. Based on the numerical and analytical results
reported above, we conjecture that the engine achieves the Carnot efficiency
(corresponding for $YT\to\infty$ to the values of bias potentials $U_{A,B}$ for 
which the efficiency is maximum, transport is dissipationless and power 
vanishes~\cite{PR694}) in the thermodynamic limit $L_\Xi\to\infty$, 
and for vanishing barriers $h_\Xi\sim 1/\sqrt{L_\Xi}\to 0$.

Of course, this conclusion holds under the condition that the dynamics in both
channels is frictionless. Two particles in a single file (say, a bullet and a rod)
do squeeze their way past each other when their relative velocity $|v_{i+1}-v_i|$
is large enough to overcome the relevant repulsive barrier $h_{\Xi}$. However, the
collisional mechanism may involve the loss of a fraction of their kinetic energy.
Accordingly, imposing pair momentum conservation, the respective momentum changes
would be $\Delta p_i=-\Delta p_{i+1}=\delta/(v_{i}-v_{i+1})$ with $\delta$ the assumed
dissipation parameter. Such a simple collisional friction model impacts the backward
current in channel $B$ more than the forward current in channel $A$; this results in 
the net suppression of the power-efficiency performance of the engine illustrated in
Fig.~\ref{fig:loops}(b).

\section{Summary and discussion}

We have exploited the phenomenon of inverse particle current  to design an autonomous
engine, which for a given temperature difference, operates without any external
time-dependent control. When operated on reverse, this engine would work as a
refrigerator. The linear-response analysis outlined above shows that the engine
performance would still be governed by $YT$ with maximum efficiency as in
Eq.~(\ref{eq:eta}) but with reversed Carnot coefficient $\eta_C=(T_L/T_R-1)^{-1}$.
Finally, due to its peculiar nature,  distinct from the conventional steady-state
engines characterized by $ZT$, our engine can be used to investigate the trade-off
of power, efficiency, and fluctuations, encompassed in thermodynamic uncertainty 
relations~\cite{TUR1, TUR2, TUR3, TUR4}.

%%%%%%%%%%%%%%%%%
\begin{figure}[!t]
\vskip-0.1cm
\includegraphics[width=8.6cm]{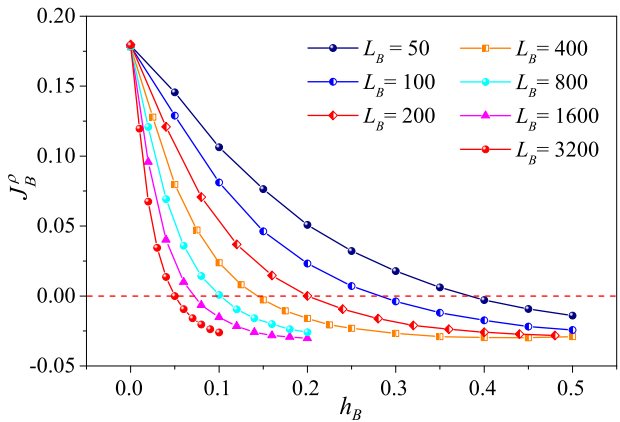}
\caption{Dependence of the current $J^\rho_B$ on the potential barrier $h_B$ for
$M_B=0.5$, $T=1$, $\mu=1.5$, $\Delta T=0.1$, and $\Delta \mu = 0$.}
\label{fig:SM1}
\end{figure}
%%%%%%%%%%%%%%%%%

\section*{ACKNOWLEDGEMENTS}

We acknowledge support by the National Natural Science Foundation of China
(Grants No. 12075198 and No. 12047501), the National Science Foundation under Grant 
No. NSF PHY-1748958, and the INFN through Project QUANTUM.

\appendix
\section{Dependence of power and efficiency on $U_A+U_B$}
\label{appA}

Given the currents [see the four equations before Eq. (\ref{eq:currents})]
and the steady-flow condition
$J^\rho_A+J^\rho_B=0$,
we obtain
$$
J^\rho_A=-J^\rho_B=\frac{1}{2}\,(J^\rho_A-J^\rho_B)
=\frac{1}{2}\,\left[
\left(
\frac{\sigma_A S_A}{L_A}-\frac{\sigma_B S_B}{L_B}
\right)\Delta T\right.
$$
\begin{equation}
\left.+
\left(
\frac{\sigma_A}{L_A}-\frac{\sigma_B}{L_B}
\right)\Delta\mu -(U_A+U_B)\right],
\end{equation}
where the last equality is obtained by substituting $J^\rho_A$ and $J^\rho_B$.
This expression shows that both
$J^\rho_A$ and $J^\rho_B$
depend on  $U_A+U_B$, rather than on $U_A$ and $U_B$,  separately.
Moreover, as
\begin{equation}
P=J^\rho_A U_A-J^\rho_B U_B= J^\rho_A(U_A+U_B),
\end{equation}
we conclude that the power depends on  $U_A+U_B$ as well.
On the other hand, by eliminating $\sigma_A (\Delta\mu -U_A)$ and
$\sigma_B (\Delta\mu -U_B)$ based on the four equations before Eq. (\ref{eq:currents}), 
we have that
\begin{equation}
\left\{
\begin{array}{l}
J^u_A=T S_A J^\rho_A + \kappa_A \Delta T/L_A,\\
J^u_B=T S_B J^\rho_B + \kappa_B \Delta T/L_B.
\end{array}
\right.
\end{equation}
It follows that $J^u_A$ and  $J^u_B$, and, in turn, the efficiency, also depend on
$U_A+U_B$ alone. In conclusion, as long as their sum is kept constant, we can vary 
$U_A$ and $U_B$ without modifying power and efficiency.

%%%%%%%%%%%%%%%%%
\begin{figure}[!b]
\vskip-0.1cm
\includegraphics[width=8.6cm]{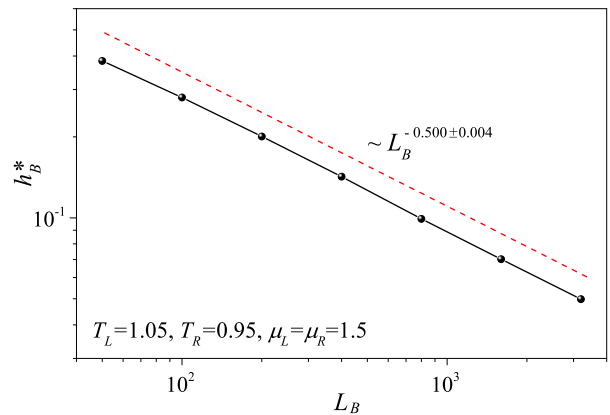}
\caption{Dependence of the critical potential barrier $h_B^\star$, obtained from the data of
Fig.~\ref{fig:SM1},  on the channel length $L_B$.}
\label{fig:SM2}
\end{figure}
%%%%%%%%%%%%%%%%%

\section{Critical value for the potential barrier}
\label{appB}

There exists a critical value $h_B^*$, such that for $h_B < h_B^*$, the bullet current in 
channel B starts flowing forward and the engine, thus, stops working. In the main text it 
was stated that $h_B^*\sim L_B^{-0.5}$. This is a very accurate numerical observation.
Figure~\ref{fig:SM1} shows the dependence of the particle current $J^\rho_B$ on the potential 
barrier $h_B$ for various channel lengths $L_B$. For a given value of $L_B$, the critical 
potential value $h_B^*$  is identified by interpolating data points and solving  
$J^\rho_B(h_B^*)=0$.  We can see from Fig.~\ref{fig:SM2} that the obtained values of the 
critical potential barrier are in excellent agreement with the scaling
$h_B^*\sim L_B^{\alpha}$ with $\alpha=-0.500\pm 0.004$.


\begin{thebibliography}{100}
\bibitem{potential}
In our simulations, we applied the potentials to both species of particles. For convenience, we used step potentials.
\end{thebibliography}
\end{document}